\documentclass[article,twocolumn,english,footinbib,tightenlines,nobibnotes,aps,prl,unsortedaddress,showpacs,superscriptaddress]{revtex4-1}
\usepackage{graphicx}
\usepackage[range-units=single,separate-uncertainty]{siunitx}
\usepackage[color links=true,linkcolor=blue,citecolor=blue,urlcolor=blue]{hyperref}

\usepackage{ulem}
\usepackage{braket}
\usepackage[usenames,dvipsnames]{color}
\usepackage{amsmath}
\usepackage{amsfonts}
\usepackage{tikz}
\usepackage{amssymb}
\usepackage{mathtools}
\usepackage{pict2e}
\usepackage{hyperref}
\usepackage{graphicx,xcolor}
\usepackage{bbding}
\usepackage{bm}
\usepackage{bbm}
\usepackage{setspace} 
\usepackage{multirow}
\usepackage{dcolumn}
\usepackage{float}
\usepackage{amsmath,mathrsfs}
\usepackage{amsthm}
\usepackage{epsfig}
\usepackage{array}
\usepackage{color}
\usepackage[caption=false]{subfig}

\usepackage{comment}

\newcolumntype{P}[1]{>{\centering\arraybackslash}p{#1}}

\def\Bk{{\bm k}}

\def\Bq{{\bm q}}
\def\BQ{{\bm Q}}

\mathchardef\sOmega="710A
\mathchardef\sGamma="7100
\mathchardef\sDelta="7101

\def\frac#1#2{{\textstyle{#1 \over #2}}}

\newcommand{\Wu}[1]{\textcolor{black}{#1}}

\begin{document}


\title{Sublattice Interference promotes Pair Density Wave order in Kagome Metals}

\author{Yi-Ming Wu}
\affiliation{Stanford Institute for Theoretical Physics, Stanford
  University, Stanford, California 94305, USA}
\author{Ronny Thomale}
\affiliation{Institute for Theoretical Physics and Astrophysics, University of W\"urzburg, D-97074 W\"urzburg, Germany}
\affiliation{Department of Physics and Quantum Centers in Diamond and Emerging Materials (QuCenDiEM) group, Indian Institute of Technology Madras, Chennai 600036, India}
\author{S. Raghu}
\affiliation{Stanford Institute for Theoretical Physics, Stanford
  University, Stanford, California 94305, USA}

\date{\today}


\begin{abstract}
Motivated by the observation of a pair density wave (PDW) in the kagome metal CsV${}_3$Sb${}_5$, we consider the fate of electrons near a p-type van Hove singularity (vHS) in the presence of local repulsive interactions.  We study the effect of such interactions on Fermi surface ``patches" at the vHS.  We show how a feature unique to the Kagome lattice known as sublattice interference crucially affects the form of the interactions among the patches.  The renormalization group (RG) flow of such interactions results in a regime where the nearest neighbor interaction $V$ exceed the onsite repulsion $U$. We identify this condition as being favorable for the formation of charge-density-wave (CDW) and PDW orders. \Wu{In the weak coupling limit, we find a complex CDW order as the leading instability, which breaks time reversal symmetry. Beyond RG, we perform a Hartree-Fock study to a $V$-only model and find the pair-density-wave order indeed sets in at some intermediate coupling.} 
\end{abstract}

\maketitle

\begin{figure}
	\includegraphics[width=7.5cm]{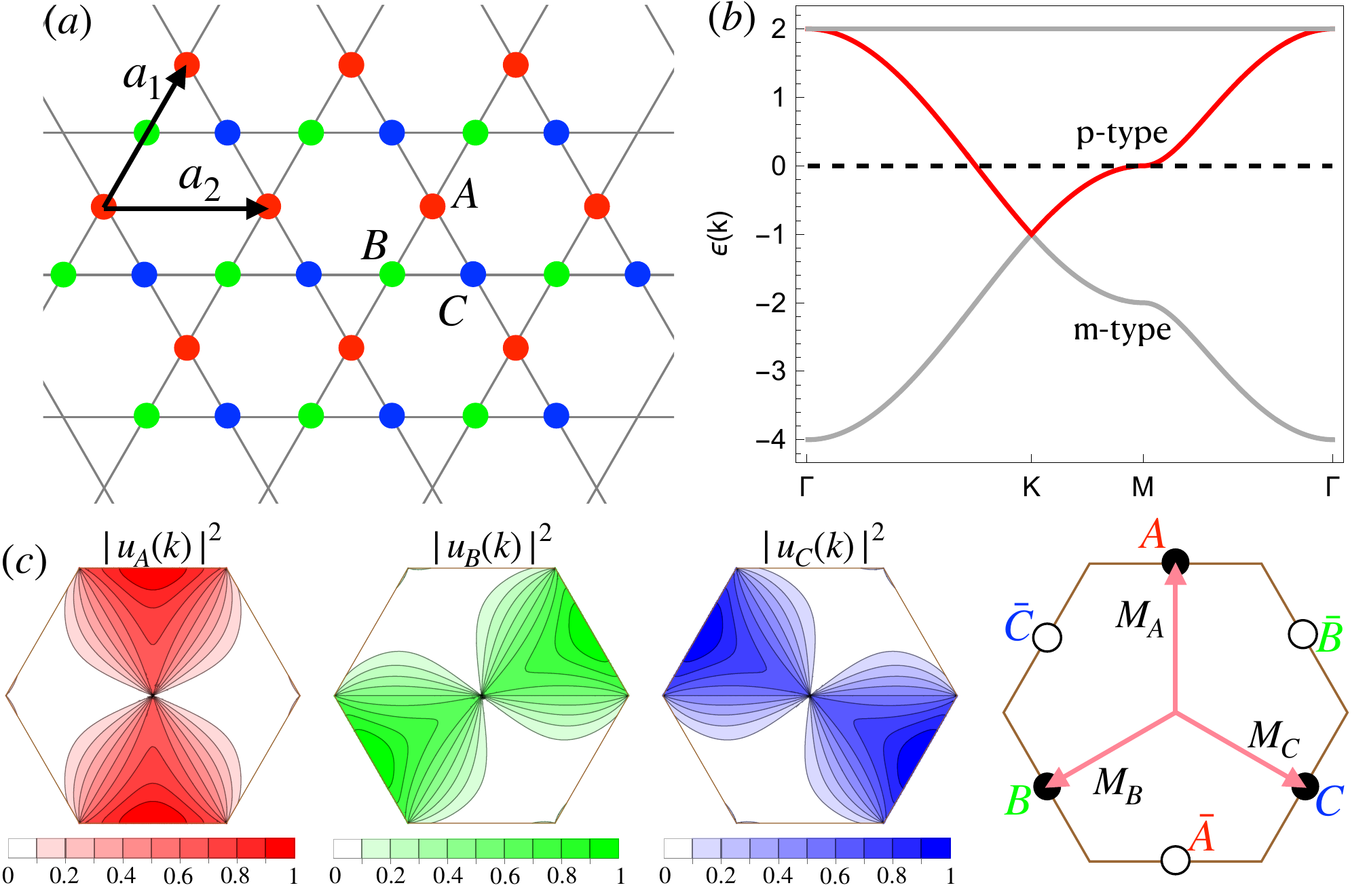}
	\caption{Sublattice interference in the Kagome lattice. (a) Each unit cell has $\alpha=\,$A,B,C different sublattice sites. (b) Band structure of the tight binding model. We focus on the middle band, highlighted with red color, which exhibits a p-type van Hove singularity. (c) For this band, the transformation matrix $u_{\alpha}(\Bk)$, defined as $c_\alpha(\Bk)=u_{\alpha}(\Bk)\psi_\alpha(\Bk)$ where $c_\alpha$ and $\psi_\alpha$ are lattice and band fermions respectively, has a modulated distribution in the Brillouin zone.
	}\label{fig:intro}
\end{figure}

\paragraph{Introduction.$-$} Spins and electrons on the Kagome lattice [Fig.\ref{fig:intro}(a)] have long been studied due to their potential for  exhibiting a panoply of exotic phases of matter.  Insulating kagome compounds, for instance, are among the most prominent candidate spin liquid materials~\cite{Han2012}, and insulating phases with non-trivial topology have also been studied on the Kagome lattice~\cite{joe-Fe3Sn2,Kang2020,Yin2020}.  With the discovery of
a family of Kagome metals AV${}_3$Sb${}_5$ (A=K,Cs,Rb), a new wave of excitement has been elicited by the prospects for intriguing density wave and superconducting ground states in these systems~\cite{PhysRevMaterials.3.094407,jiangii,neupert-review}.

Among the more fascinating observed phenomena in these kagome metals include new bragg-like peaks inside the superconducting phase~\cite{roton-pdw}, a hallmark of an exotic superconducting order known as a pair density wave (PDW)~\cite{doi:10.1146/annurev-conmatphys-031119-050711,Fradkin2015,Himeda2002,Yang_2009,Raczkowski2007,Berg2007,Capello2008,PALee2014,Wang2015a,Wang2015b,Edkins2019,Wang2018,Zhengzhi,tTMD,Shaffer1,Shaffer2}
PDWs are superconductors with an order parameter that varies periodically in space.  Additionally, fascinating effects have been discovered in the superconducting fluctuation spectrum, including nearly condensed excited states with charge 4e, 6e superconducting fluctuations~\cite{ge2022}.  These observations call for a greater theoretical scrutiny, and invite us to  make predictions for electronic phases of kagome metals.  

In this Letter, motivated by these recent developments, we address the issue of whether the PDW superconducting phase can in principle arise on the Kagome lattice.  Since such superconductivity requires analysis of the intermediate coupling problem, robust pairing mechanisms for PDW formation have only recently been uncovered~\cite{WuPDWlargeN}.  One key requirement for PDW order is the presence of strong repulsive interactions with somewhat suppressed onsite interactions.  While this is rather unusual in most solids, we show here that this requirement is met when the chemical potential crosses a so-called p-type (for ``pure") van Hove singularity (vHS)~\cite{PhysRevB.86.121105,PhysRevLett.127.177001,kang,hu2021rich} [see Fig.\ref{fig:intro}(b)].  
In an important theoretical study, one of us showed~\cite{PhysRevB.86.121105} that precisely at such a p-type vHS, there is the phenomena of sublattice interference (SI)  - where each of the 3 distinct sublattices has non-zero support only on {\it one} of the 3 distinct vH points.  
To show this SI crucially determines the low energy effective interactions in the system,
we construct a renormalization group (RG) theory 
based on 
the p-type vHS and show that the onsite repulsion runs towards weak coupling, while nearest neighbor interactions grow under the RG.
This peculiarity results in a rich phase diagram including time-reversal symmetry breaking charge-density-wave (CDW) orders and various uniform superconductivity in the weak coupling limit; while in the strong coupling regime it
makes the Kagome system poised to exhibit the PDW order.
 While more experiments are needed to characterize precisely the normal state in this system, our analysis already establishes that the kagome motif once again provides us with an avenue towards an exotic phase of matter, in this case the PDW superconductor.

{\it SI and the triviality of Hubbard $U$.$-$} 
To determine the fate of effective interactions at p-type van Hove filling, 
we utilize parquet RG methods and restrict attention to Fermi surface ``patches" in the neighborhood of the distinct van Hove points~\cite{Dzyaloshinskii1987,Schulz1987} (similar RG analysis with two-fold of van Hove singularities is considered in Ref.\cite{Scammell2023}).  Defining the  fermion destruction operators in  patch $\alpha$ as $\psi_\alpha$, we encode real-space 4-fermion interactions as intra- and interpatch couplings after Fourier transformation.   
In a crystal with time-reversal and/or inversion symmetry, such interactions  take the form $H_I= g_1\psi_{\alpha}^\dagger\psi_{\beta}^\dagger\psi_{\alpha}\psi_{\beta}+g_2\psi_{\alpha}^\dagger\psi_{\beta}^\dagger\psi_{\beta}\psi_{\alpha}+g_3\psi_{\alpha}^\dagger\psi_{\alpha}^\dagger\psi_{\beta}\psi_{\beta}+g_4\psi_{\alpha}^\dagger\psi_{\alpha}^\dagger\psi_{\alpha}\psi_{\alpha}$ where $\alpha\neq\beta$ and the momentum summation and a spin configuration $\sigma\sigma'\sigma'\sigma$ is assumed. Patch models have been applied to the square lattice\cite{Furukawa}, the honeycomb lattice~\cite{Nandkishore2012,PhysRevB.89.144501} and moiré systems\cite{Isobe2018,PhysRevB.100.085136,Zhengzhi} to capture interaction-driven electronic orders. In both cases, {\it all} the $g_i$'s are set by the largest Hubbard onsite interaction $H_U=U\sum_i n_{i\uparrow}n_{i\downarrow}$.

However,
adopting the same strategy on the Kagome lattice near a p-type vHS, one 
finds that 
the Hubbard interaction U 
contributes only to the $g_4$.
This is due to SI: Near a  p-type vHS, each of the 3 sublattices (A,B,C) has non-zero support only on {\it one} of the 3 distinct van Hove points $(\bm M_{\alpha}, \alpha = A,B,C)$ which is visible through the sublattice weight in Fig.\ref{fig:intro}(c). The SI does not result from fine tuning, as it exists even if longer range hoppings are considered~\cite{supp}.
Thus, when only $H_U$ is present for electrons near a p-type vHS,  $g_1=g_2=g_3=0$, and $g_4=U/t$.
With this choice of bare couplings,  $g_4$ weakens under RG flow, eventually reaching a trivial fixed point with $g_4^*=0$~\cite{supp}. Thus, due to SI,  the repulsive Hubbard U is irrelevant in the RG sense near the p-type vHS of the Kagome lattice \footnote{For comparison, one can also consider the alternative kagome van Hove filling which is of m-type [see Fig.\ref{fig:intro}(b)]. There, each $\psi_\alpha$ is a superposition of two distinct lattice fermions and it can be directly verified that the Hubbard $U$ then has equal contributions to all $g_i$.}. 

\begin{figure}
	\includegraphics[width=8.5cm]{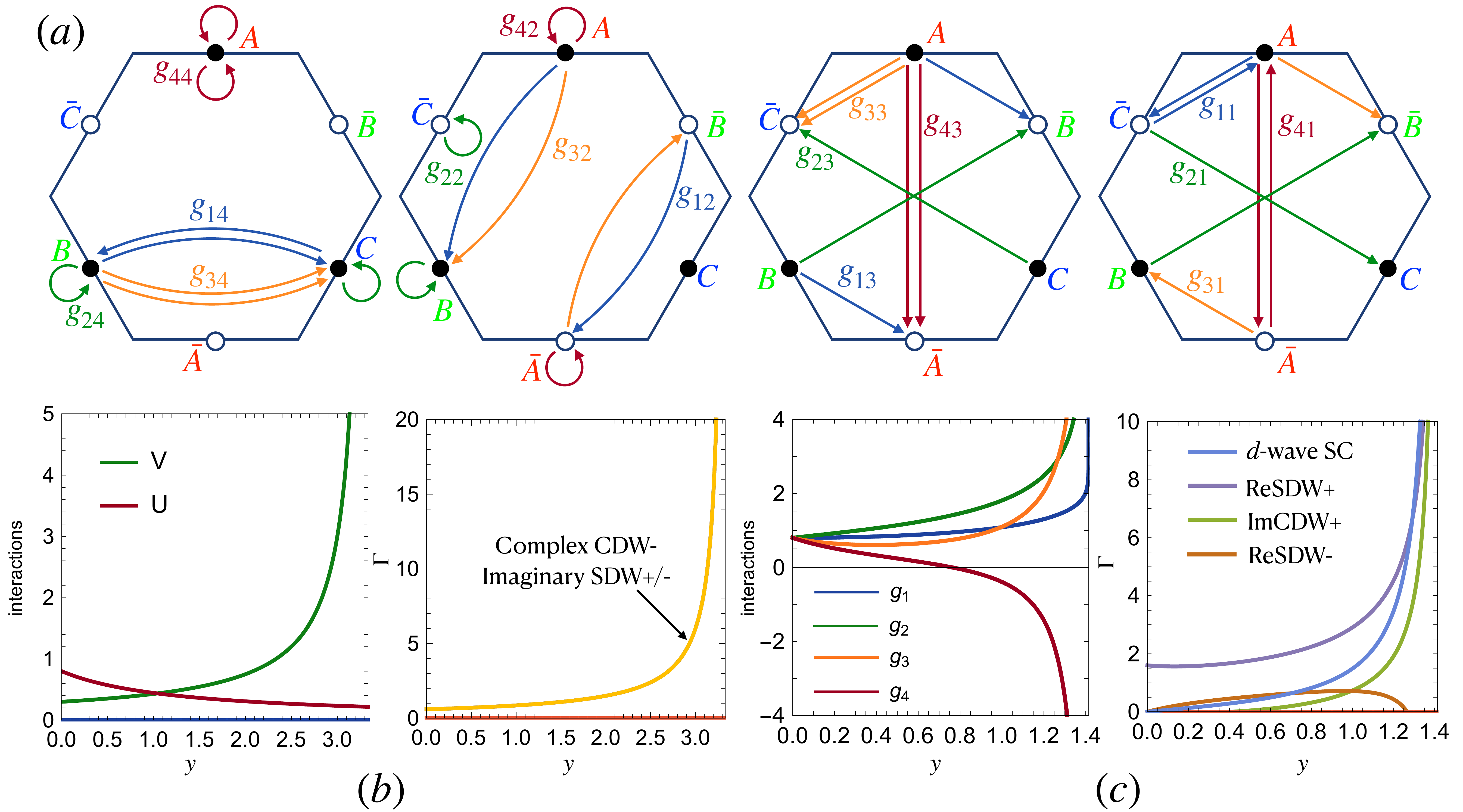}
	\caption{(a) All symmetry allowed interactions in the six patch model.
	\Wu{(b) RG flow for the interactions and the order parameter vertex $\Gamma$ with the SI effect. We set $U(0)=0.8t$ and $V(0)=0.3t$ and starting from some intermediate energy scale we have $V\gg U$. The resulting weak coupling instabilities are degenerate complex CDW$_-$ and imaginary SDW$_{+/-}$. (c) The same RG flow but without SI effect. In this case all the $g_{ij,0}$ are set by the largest Hubbard $U$,  and there does not exist a constant map  $U,V$ and $g_{ij}(y)$. The leading weak coupling instability is the $d$-wave SC\cite{Nandkishore2012,PhysRevB.89.144501}.}}\label{fig:g}
\end{figure}

{\it Extended interactions and the 6-patch theory.$-$} Due to the apparent irrelevance of $U$, it is necessary to include at least nearest neighbor interactions.
The most important such term 
 is the nearest neighbor density-density repulsion $H_V=V\sum_{\braket{ij}}n_in_j.$
With both $H_U$ and $H_V$, the  3-patch model described above is no longer adequate 
since nearest neighbor interactions contain momentum dependence, which differentiates the patches at $\pm \bm{M}_\alpha$. Therefore, we need to consider the full 6-patch theory, for which there are 16 different symmetry allowed interactions in total, as shown in Fig.\ref{fig:g}(a). 
Here we adopt the convention used in Ref.\cite{Isobe2018} and define the patch fermions as $\psi_{\alpha+}=\psi(\Bk\to\bm{M}_\alpha)$ and $\psi_{\alpha-}=\psi(\Bk\to-\bm{M}_\alpha)$.  The interactions in this six patch model can be written as 
\begin{equation}
H_I=\sum_{i,j=1}^4\sum_{\left\{ \alpha_j, \tau_j \right\}}g_{ij}\psi^\dagger_{\alpha_1\tau_1}\psi^\dagger_{\alpha_2\tau_2}\psi_{\alpha_3\tau_3}\psi_{\alpha_4\tau_4},
\end{equation} 
with $g_{ij}$ being the dimensionless interaction strengths.  
Momentum conservation constrains the indices as follows: the patch indices satisfy $\alpha_1=\alpha_3\neq\alpha_2=\alpha_4$ for $i=1$, $\alpha_1=\alpha_4\neq\alpha_2=\alpha_3$ for $i=2$, $\alpha_1=\alpha_2\neq\alpha_3=\alpha_4$ for $i=3$, and $\alpha_1=\alpha_2=\alpha_3=\alpha_4$ for $i=4$. The `valley' indices $\tau_i=\pm$, labeling whether the patch is at $\bm{M}_{\alpha}$ or $-\bm{M}_{\alpha}$, obey the same rule associated with $j$. That is, $\tau_1=\tau_3\neq\tau_2=\tau_4$ for $j=1$ et cetera. 

Once again, SI crucially influences the initial conditions of the RG flows.  A straightforward calculation~\cite{supp} shows that  $U$ contributes only to $g_{4j}$ - a direct generalization of the 3-patch theory - while $V$ only contributes to $g_{2j}$: 
\begin{equation}
  g_{4j,0}=\frac{U}{t}; g_{22,0}=g_{24,0}=-g_{21,0} = - g_{23,0} = \frac{2V}{t}.\label{eq:SI}
\end{equation}
The subscript $0$ above denotes bare interactions before running RG. From microscopics, it is natural to expect that $U$ (and therefore $g_{4j,0}$) should be the largest.

\Wu{{\it CDW  at weak coupling.$-$} We first investigate the weak coupling limit. In this case the fate of the fermions can be described by the one loop RG equations of $g_{ij}$, for which we keep the most divergent ``log squared" terms in perturbation theory, i.e. the particle-hole bubble $\Pi_{ph}^{(0)}$ at momenta $\bm{M}_\alpha$ and the particle-particle bubble $\Pi_{pp}^{(0)}$ at zero momentum, and set the running parameter as the latter: $y=\Pi_{pp}^{(0)}(0)$~\cite{Isobe2018,supp}. With this convention, the RG equations for the interactions take the form $dg_{ij}/dy= g_{mn}R_{mn,kl}g_{kl}$. It's easy to see that close to the critical value $y_c$ all $g_{ij}$ behaves in a similar manner, namely $g_{ij}=G_{ij}/(y_c-y)$. If some $G_{ij}$ is nonzero, the corresponding $g_{ij}$ thus diverges at $y_c$.
A typical order parameter flows as $d\Delta_i/dy=d\Gamma_i\Delta_i$ under renormalization, where $d=1$ for superconducting orders and $d=d_1\leq\frac{1}{2}$ for density wave orders, and $\Gamma_i$ is certain linear combination of $g_{ij}$\cite{supp}.  From this we can integrate to obtain the behavior of the corresponding susceptibility $\chi_i$ which scales as $\chi_i\sim(y_c-y)^{\gamma_i}$. The leading instabilities will be those with the most negative $\gamma_i$, and accordingly the most divergent $\Gamma_i$. We mainly consider the following weak coupling instabilities,
\begin{equation}
	\begin{aligned}
		\Delta_{\text{CDW}_\pm}&=\braket{\psi_{\alpha+}^\dagger\psi_{\beta+}\pm\psi_{\alpha-}^\dagger\psi_{\beta-}},~\Delta_{\text{SC}}=\braket{f_\alpha\psi_{\alpha+}\psi_{\alpha-}},\\
		\Delta_{\text{SDW}_\pm}&=\braket{\psi_{\alpha+}^\dagger\bm{\sigma}\psi_{\beta+}\pm\psi_{\alpha-}^\dagger\bm{\sigma}\psi_{\beta-}}.
	\end{aligned}
\end{equation}
Note that the form factor $f_{\alpha}=\pm1$ in $\Delta_{\text{SC}}$ determines the pairing symmetry. For the density wave orders, their real and imaginary parts flow differently, so depending on whether their imaginary parts are zero or not, the system could either preserve or break the time reversal symmetry.}

\Wu{In Fig.\ref{fig:g}(b) we show the RG results in the presence of SI, i.e. the initial values of $g_{ij}$ are set by Eq.\eqref{eq:SI}. Interestingly, we find that this map between lattice interactions and $g_{ij}$ persists for all $y<y_c$, and it is this constant map that enables us to extract the RG flows for $U$ and $V$. We see the $U$ decays as before, while $V$ increases. The leading instability in this case is the degenerate complex $\Delta_{\text{CDW}_-}$ and imaginary $\Delta_{\text{SDW}_\pm}$. The presence of imaginary parts for these density wave orders indicate the time reversal symmetry is broken\cite{Jiang2021,yu2021evidence,Mielke2022,FENG20211384,PhysRevB.104.045122,PhysRevLett.129.167001,PhysRevB.104.035142},, which is due to the SI effect. For comparison, we also show in Fig.\ref{fig:g}(c) the RG analysis for the model but with no SI effect. In this case, all $g_{ij}$ are set by the largest interaction $U$, and the resulting weak coupling instability is the $d$-wave uniform superconductivity, consistent with Ref.\cite{Nandkishore2012,PhysRevB.89.144501}.}

{\it \Wu{PDW at intermediate coupling}.--}Having established the way different interactions get renormalized in the presence of the SI effect, we now discuss how the PDW order can emerge at some intermediate energy scale. 
\Wu{Based on the observation discussed above [see Fig.\ref{fig:g}(b) for example], we consider an effective model where only a large $V$ is kept. Model similar this has also been studied recently in Ref.\cite{Dong_2023}. We consider a sufficiently large $V$, and perform a Hartree-Fock mean-field study\cite{PhysRevB.104.075150,PhysRevLett.124.187601,PhysRevB.104.214403,PhysRevB.102.035136,PhysRevB.102.201104} for the corresponding orders\footnote{Justification of the Hartree-Fock analysis can be achieved by generalizing the model to a large-$N$ theory, where fluctuations around the saddle point are suppressed by $1/N$ and the Hartree-Fock mean field equations become asymptotically exact. See \cite{doi:10.1146/annurev-conmatphys-031620-102024} and references therein for examples.}. }
Without loss of generality, we discuss the $AB$ bond only, as the other bonds follow via $C_3$ rotation. 
On the AB bond, the relevant interaction is $2V\cos(\Bq\cdot\bm{\alpha})\psi_{A\sigma}^\dagger(\Bk)\psi_{B\sigma'}^\dagger(\Bk')\psi_{B\sigma'}(\Bk'+\Bq)\psi_{A\sigma}(\Bk-\Bq)$ where $\bm{\alpha}=\bm{a}_1/2$ and $\bm{a}_1$ is the vector connecting two adjacent $A$ and $B$ sites [see Fig.\ref{fig:intro} (a)]. Whether this is repulsive or attractive depends on the momentum transfer $\Bq$. In the Cooper channel, we can write it as $2V\cos[(\Bk+\Bk'-\Bq)\cdot\bm{\alpha}]\psi_{A\sigma}^\dagger(\Bk)\psi_{B\sigma'}^\dagger(-\Bk+\Bq)\psi_{B\sigma'}(\Bk')\psi_{A\sigma}(-\Bk'+\Bq)$, which is attractive when $\cos[(\Bk+\Bk'-\Bq)\cdot\bm{\alpha}]<0$. In the patch model this condition is met when  $\Bk$ is around $\bm{M}_A$, $-\Bk+\Bq$ is around $-\bm{M}_B$, and $\Bk'$ is around $-\bm{M}_A$. The interaction then becomes $-2V\psi_{A}^\dagger\psi_{\bar{B}}^\dagger\psi_B\psi_{\bar{A}}$. When $V$ is large, we can use
\begin{equation}	\psi_{A}^\dagger\psi_{\bar{B}}^\dagger\psi_B\psi_{\bar{A}}\approx\braket{\psi_{A}^\dagger\psi_{\bar{B}}^\dagger}\psi_B\psi_{\bar{A}}+\psi_{A}^\dagger\psi_{\bar{B}}^\dagger\braket{\psi_B\psi_{\bar{A}}}\label{eq:PDWmeanfield}
\end{equation}
for mean field analysis, and the gap function, defined as $\Delta_{\BQ}\sim\braket{\psi_B\psi_{\bar{A}}}$, is apparently a PDW order with momentum $\bm{M}_C$. {We note that, due to the presence of SI, the uniform superconductivity is not a competing order \Wu{in the large $V$ model}, since such order couples to,  e.g. $\psi^\dagger_A\psi_A^\dagger$, which can not be obtained by decomposing Eq.\eqref{eq:PDWmeanfield}.

Similarly, the onsite CDW does not arise since this order parameter couples to, e.g. $\psi^\dagger_A\psi_A$, and effectively becomes the chemical potential. However, the bond charge density wave (bond CDW) order can indeed arise when $V$ becomes strong~\cite{PhysRevLett.110.126405}, and compete with PDW. This occurs also when
$\cos(\Bq\cdot\bm{\alpha})<0$
and the density interaction becomes $-2V\psi_{A}^\dagger\psi_{\bar{B}}^\dagger\psi_B\psi_{\bar{A}}$ with $V>0$. We can, however, consider 
\begin{equation}	-\psi_{A}^\dagger\psi_{\bar{B}}^\dagger\psi_B\psi_{\bar{A}}\approx\braket{\psi_{A}^\dagger\psi_B}\psi_{\bar{B}}^\dagger\psi_{\bar{A}}+\psi_{A}^\dagger\psi_B\braket{\psi_{\bar{B}}^\dagger\psi_{\bar{A}}}\label{eq:CDW}
\end{equation}
to absorb the minus sign and  manifestly show it contains strong repulsion in the particle-hole channel. Moreover, the spin indices on the left hand side of Eq.~\ref{eq:CDW} can be included explicitly, namely $-\psi_{A\alpha}^\dagger\psi_{\bar{B}\beta}^\dagger\psi_{B\gamma}\psi_{\bar{A}\delta}\delta_{\alpha\delta}\delta_{\beta\gamma}$, and using the $SU(2)$ identity $\delta_{\alpha\delta}\delta_{\beta\gamma}=(\sigma_{\alpha\gamma}\delta_{\beta\delta}+\bm{\sigma}_{\alpha\gamma}\bm{\sigma}_{\beta\delta})/2$, it is easy to see that the order parameter $\Delta_{\bm{Q}}=\braket{\psi_{A}^\dagger\psi_B}$ represents a bond CDW/SDW order which are directly related to those density waves that arise in the weak coupling limit. In fact, the degeneracy is an artifact of the oversimplification of the patch model. Considering that the SDW order has to break a global $SU(2)$ symmetry, and upon taking the whole Fermi surface into account, we assume for the time being that the CDW order is more likely to occur than SDW. In the following we thus constrain ourselves to CDW order competing with PDW at large $V$.

To inspect the competition between PDW and CDW at large $V>0$, we can calculate the susceptibilities for both of the two orders in random-phase-approximation (RPA). The result is $\chi_{PDW}=\Pi_{pp}^0(\BQ)/[1-2V\Pi_{pp}^0(\BQ)]$ and $\chi_{CDW}=\Pi_{ph}^0/[1-2V\Pi_{ph}^0(\BQ)]$. Here the bare susceptibilities $\Pi^0_{pp}>0$ and $\Pi_{ph}>0$ are obtained using free fermion propagators. The ordering condition is given when the denominators vanish.
The problem is now reduced to comparing the strength of the bare $\Pi_{pp}^{(0)}$ and $\Pi_{ph}^{(0)}$. Within the patch model, the particle-particle and particle-hole  bubbles can be obtained in a straightforward way, and the results are\footnote{The apparent dependence on the artificial energy scale $\Lambda$ is unphysical, and there should be contributions from outside the patches which eventually add up to make the final results larger and independent of $\Lambda$.}
\begin{equation}
	\begin{aligned}
		\Pi_{pp}^{(0)}(\BQ)&=\frac{1}{4\sqrt{3}\pi^2t}\ln\frac{\Lambda}{\max\{T,|\mu|\}}\\
		\Pi_{ph}^{(0)}(\BQ)&=\frac{1}{8\sqrt{3}\pi^2t}\ln\frac{\Lambda}{\max\{T,|\mu|\}}\ln\frac{\Lambda}{\max\{T,\mu,|t'|\}},
	\end{aligned}\label{eq:patchPi}
\end{equation}
where $\mu=0$ at the van Hove filling, and $t'$ is the next nearest neighbor hopping which can spoil the Fermi surface nesting in the particle-hole channel. In Fig.~\ref{fig:PI}(a) we compare both $\Pi_{pp}^{(0)}$ and $\Pi_{ph}^{(0)}$ as a function of temperature $T$ at $\mu-0$. We see that the $\ln^2$ divergence of $\Pi_{ph}^{(0)}$ becomes dominant at small $T$ for the case of perfect nesting ($t'=0$), while at larger $T$, $\Pi_{pp}^{(0)}$ is larger even with perfect nesting. The consequence is that, in the weak coupling limit, one has to go to small $T$ to see the instability, where CDW wins over PDW. For large $V$, however, a relatively smaller $\Pi_{pp}^{(0)}$ or $\Pi_{ph}^{(0)}$ is enough to induce the instability, which could be in the regime where $\Pi_{pp}^{(0)}>\Pi_{ph}^{(0)}$ and PDW is the leading instability. Note that Eq.\eqref{eq:patchPi} is only a crude estimation about the relative strength between $\Pi_{pp}^{(0)}(\BQ)$ and $\Pi_{ph}^{(0)}(\BQ)$. \Wu{Away from the van Hove filling ($\mu\neq0$), we expect that both bubbles are reduced, and therefore one needs to reach smaller $T$ in order to reach the instability. This results in the schematic picture we show as the inset of Fig.\ref{fig:g}(a). }

\begin{figure}
	\includegraphics[width=8.cm]{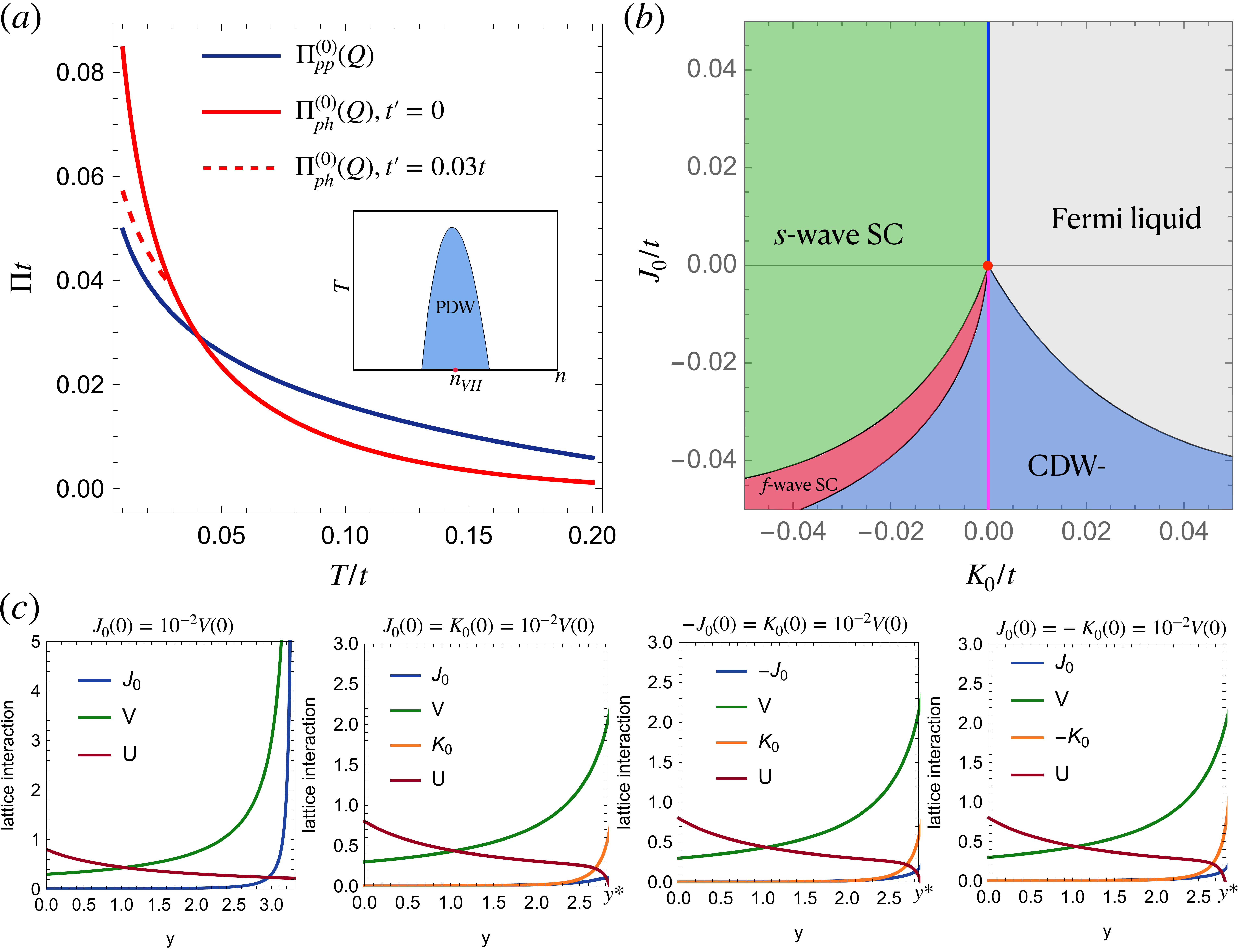}
	\caption{(a)Comparison between $\Pi_{pp}^{(0)}(\BQ)$ and $\Pi_{ph}^{(0)}(\BQ)$ as a function of temperature $T$. \Wu{The fact that $\Pi_{pp}^{(0)}(\BQ)>\Pi_{ph}^{(0)}(\BQ)$ at larger $T$ indicates that PDW order may win over CDW order when interaction is sufficiently strong. (b) The weak coupling instabilities (at perfect nesting) when other interactions such as $J_0$ and $K_0$ are taken into account. (c) RG flows for the lattice interactions for $U(0)=0.8t$ and $V(0)=0.3t$ and for various $J_0$ and $K_0$. In all cases there is a wide range in energy scale where $V$ is the largest.}}\label{fig:PI}
\end{figure}

{\it Other interactions.} We note that other than $U$ and $V$, further lattice interaction might be worth analyzing. For instance, we can consider the bond singlet pair hopping term, for which, again taking $AB$ bond as an example, the on-bond contribution is $J_0\sum_{i}P^\dagger_{AB}(\bm{r}_i)P_{AB}(\bm{r}_i)$, 
where the pair operator at position $\bm{r}_i$ is given by $P^\dagger_{AB}(\bm{r}_i)=c_{A\uparrow}^\dagger(\bm{r}_i+\bm{\alpha})c^\dagger_{B\downarrow}(\bm{r}_i)-c_{A\downarrow}^\dagger(\bm{r}_i+\bm{\alpha})c^\dagger_{B\uparrow}(\bm{r}_i)$ and $\bm{\alpha}=\bm{a}_1/2$ is the vector connecting two adjacent $A$ and $B$ sites. In fact, the $J_0$ term just reads as the exchange interaction. 
{The site pair hopping term such as $K_0\sum_{\langle ij\rangle}P^\dagger_iP_j$ with $P_i^\dagger=c_{i\uparrow}^\dagger c_{i\downarrow}^\dagger$ is also possible. }
With the SI effect, it is easy to realize that $J_0$ 
contributes to both $g_{1j}$ and $g_{2j}$. In particular, $g_{11}(0)=g_{14}(0)=-g_{12}(0)=-g_{13}(0)=2J_0$, and $-g_{21}(0)=-g_{23}(0)=g_{22}(0)=g_{24}(0)=2V+2J_0$. {The $K_0$ term only contributes to $g_{3j}$: $g_{31}(0)=g_{32}(0)=-g_{33}(0)=-g_{34}(0)=2K_0$}. \Wu{We expect that $J_0$ and $K_0$ are of the same order, and both are orders of magnitudes  smaller than $U$ and $V$\cite{doi:10.1098/rspa.1963.0204,PhysRevLett.58.1899,PhysRevLett.60.71,PhysRevLett.60.72}. In the presence of $J_0$ and $K_0$, both of them increase under RG, leading to different orders in the weak coupling limit which we present in Fig.\ref{fig:PI}(b). These include $s$- and $f$-wave uniform SC, the real CDW order with odd parity (CDW$_-$). Interestingly, if we keep $K_0=0$ there is still a constant map between $U, V, J_0$ and $g_{ij}$ for all $y<y_c$, and the weak coupling instabilities are degenerate imaginary $\Delta_{\text{SDW}_\pm}$ for $J_0>0$, and a complex $\Delta_{\text{CDW}_-}$ for $J_0<0$ [see the blue and purple line in Fig.\ref{fig:PI}(b)]. A finite $K_0$ spoils the constant map between lattice interactions and $g_{ij}$ at $y$ near $y_c$, but at some intermediate $y^*<y_c$ we can still approximately determine the flow of $U,V,J_0$ and $K_0$. Fig.\ref{fig:PI}(c) (see also in \cite{supp}) shows that $V$ can still be the largest at intermediate energy scale for various cases, which justifies the applicability of our effective $V$ model discussed above. We close by noting the possibility that when a sizable $K_0$ is present initially, $K_0$ could become comparable to $V$ even at $y<y^*$. Since $K_0$ is related to the formation of uniform SC, one needs to study the competition between uniform SC and PDW in this case.  }


\paragraph{Discussion.-} We have focused our analysis entirely on the case of electrons subject to instantaneous repulsive interactions on the Kagome lattice near p-type van Hove filling. At the current level of understanding, it is not yet settled to which extent the kagome metals AV$_3$Sb$_5$ (A=K,Cs,Rb) are faithfully represented by this simplified model. Previous attempts to improve on the microscopic rigor by taking into account the multi-orbital nature at the Fermi level~\cite{PhysRevLett.127.177001,PhysRevLett.127.217601}, combined with the relevance of phononic contributions~\cite{PhysRevLett.127.046401}, suggest the need for further analysis to quantitatively approach the experimental setup in AV$_3$Sb$_5$, and further experiments involving Kagome metals are necessary to validate the nature of the CDW observed at scales as high as $T \sim 90K$. In particular, such order would of course affect the reconstructed bands out of which the superconductivity ultimately develops.   A coherent theory of such orders, including the effects of phonons, will only be feasible once the precise nature of the CDW order in this system is well-characterized.  

Our analysis does, however, make a rare {\it microscopically founded} statement about the realization of PDW in a physically sensible model, specifically in a kagome metal nearby a p-type vHS with local and nearest neighbor Coulomb repulsion. This is because SI crucially affects the renormalization of repulsive interactions in Kagome metals, and thus the extended repulsive forces are enhanced relative to the onsite Hubbard repulsion.  As a result, there is an increased tendency towards PDW and bond CDW order, which we identified without any need for a potentially biased mean field analysis.  Further input from experiment is needed at this stage to bridge the ideas described in this paper with the thus far concluced phase diagram of Kagome metals.



\paragraph{Acknowledgement.} We sincerely thank Mengxing Ye for useful discussions, and J.~Beyer, M.~D\"urrnagel, T.~M\"uller,  J.~Potten, and  T.~Schwemmer for ongoing collaborations on related topics.  Y.-M. W. acknowledges the Gordon and Betty Moore Foundation’s EPiQS Initiative through GBMF8686 for support at Stanford University.R.T. acknowledges support from the Deutsche Forschungsgemeinschaft (DFG, German Research Foundation) through QUAST FOR 5249-449872909 (Project P3), through Project-ID 258499086-SFB 1170, and from the W\"urzburg-Dresden Cluster of Excellence on Complexity and Topology in Quantum Matter – ct.qmat Project-ID 390858490-EXC 2147. S.R. is supported by the Department of Energy, Office of Basic Energy Sciences, Division of Materials Sciences and Engineering, under Contract No. DE-AC02-76SF00515.


\bibliographystyle{prsty}
\bibliography{av3sb5}

\newpage
\begin{widetext}

	\begin{center}
  \textbf{\large  ONLINE SUPPORTING MATERIAL
  \\[.2cm] Sublattice Interference promotes Pair Density Wave in Kagome Metals}
\\[0.2cm]
Yi-Ming Wu,$^{1}$, Ronny Thomale$^{2}$ and S. Raghu$^{1}$
\\[0.2cm]
{\small \it $^{1}$ Stanford Institute for Theoretical Physics, Stanford University, Stanford, California 94305, USA}

{\small \it $^{2}$ Institute for Theoretical Physics and Astrophysics, University of W\"urzburg, D-97074 W\"urzburg, Germany}

  \vspace{0.4cm}
  \parbox{0.85\textwidth}{In this Supplemental Material we show i) the sublattice interference is robust against the inclusion of long range hoppings, ii) how the sublattice interference sets the initial values for different patch interactions and iii) the detailed calculation for the bare PDW and CDW susceptibilities.} 
\end{center}

\section{Robustness of Sublattice inteference} 
\label{sec:robustness_of_sublattice_inteference}

The tight binding model for the Kagome lattice including the nearest neighbor (NN) hopping $t$, the next nearest neighbor (NNN) hopping $t'$ and the next-next nearest neighbor (NNNN) hopping $t''$ is given by 
\begin{equation}
	H_0=\sum_{\Bk}(c_{A}^\dagger(\Bk),c_{B}^\dagger(\Bk),c_{C}^\dagger(\Bk))\mathcal{H}(\Bk)
	\begin{pmatrix}
		c_A(\Bk)\\
		c_B(\Bk)\\
		c_C(\Bk)\\
	\end{pmatrix}
\end{equation}
where 
\begin{equation}
	\mathcal{H}(\Bk)=\begin{pmatrix}
		2t''\sum_{i=1}^3\cos(\Bk\cdot\bm{a}_i) &2t\cos(\Bk\cdot\bm{a}_1/2)+2t'\cos(\Bk\cdot\bm{A}_1) & 2t\cos(\Bk\cdot\bm{a}_3/2)+2t'\cos(\Bk\cdot\bm{A}_3)\\
		2t\cos(\Bk\cdot\bm{a}_1/2)+2t'\cos(\Bk\cdot\bm{A}_1)& 2t''\sum_{i=1}^3\cos(\Bk\cdot\bm{a}_i) & 2t\cos(\Bk\cdot\bm{a}_2/2)+2t'\cos(\Bk\cdot\bm{A}_2) \\
		2t\cos(\Bk\cdot\bm{a}_3/2)+2t'\cos(\Bk\cdot\bm{A}_3)& 2t\cos(\Bk\cdot\bm{a}_2/2)+2t'\cos(\Bk\cdot\bm{A}_2) & 2t''\sum_{i=1}^3\cos(\Bk\cdot\bm{a}_i)\\
	\end{pmatrix}
\end{equation}
and 
\begin{equation}
	\begin{aligned}
		\bm{a}_1&=\frac{1}{2}(1,\sqrt{3}), ~\bm{a}_2=(1,0), ~\bm{a}_3=\frac{1}{2}(-1,\sqrt{3})\\
		\bm{A}_1&=\frac{1}{4}(-3,\sqrt{3}), ~\bm{A}_2=\frac{1}{2}(0,\sqrt{3}), ~\bm{A}_3=\frac{1}{4}(3,\sqrt{3})
	\end{aligned}
\end{equation}

The Hamiltonian can be diagonalized as $H_0=\sum_{n,\Bk}\psi_n^\dagger(\Bk)\psi_n(\Bk)$ by using the following linear transformation between the band fermions and the lattice fermions:
\begin{equation}
  c_s(k)=\sum_nu_{sn}(k)\psi_n(k)\label{eq:transform2}
\end{equation}
where $s\in(A,B,C)$ and $n\in(1,2,3)$ is the band index. If we focus on the middle band of the kagome lattice where the p-type VHS locates, we can omit the band index $n$. The weight $u_{s}(\Bk)$ shows strong sublattice interference, i.e. the distribution in momentum space strongly depends on the sublattice index $s$. For clarity, in Fig.\ref{fig:SI_check} we show the configuration of $|u_A(\Bk)|^2$ for different situations with and without including longer range hopping amplitudes. The results show that even in the presence of $t'$ and $t''$, $|u_A(\Bk)|^2$ changes little. 

If we consider only three patches in the small vicinity of these VHS points ($\bm{M}_A,\bm{M}_B,\bm{M}_C$), we find that $u_s(\Bk)$ satisfies
\begin{equation}
 \begin{aligned}
   \text{if} ~\Bk\in \bm{M}_A, ~u_A(\Bk)=1,~ u_B(\Bk)=u_C(\Bk)=0, \\
   \text{if} ~\Bk\in \bm{M}_B, ~u_B(\Bk)=1,~ u_A(\Bk)=u_C(\Bk)=0, \\
   \text{if} ~\Bk\in \bm{M}_C, ~u_C(\Bk)=1,~ u_A(\Bk)=u_B(\Bk)=0.
 \end{aligned}\label{eq:interference}
\end{equation}
These results are immune to the presence of a nonzero $t'$ or $t''$, thus the sublattice interference is a robust phenomenon.

\begin{figure}
	\includegraphics[width=12cm]{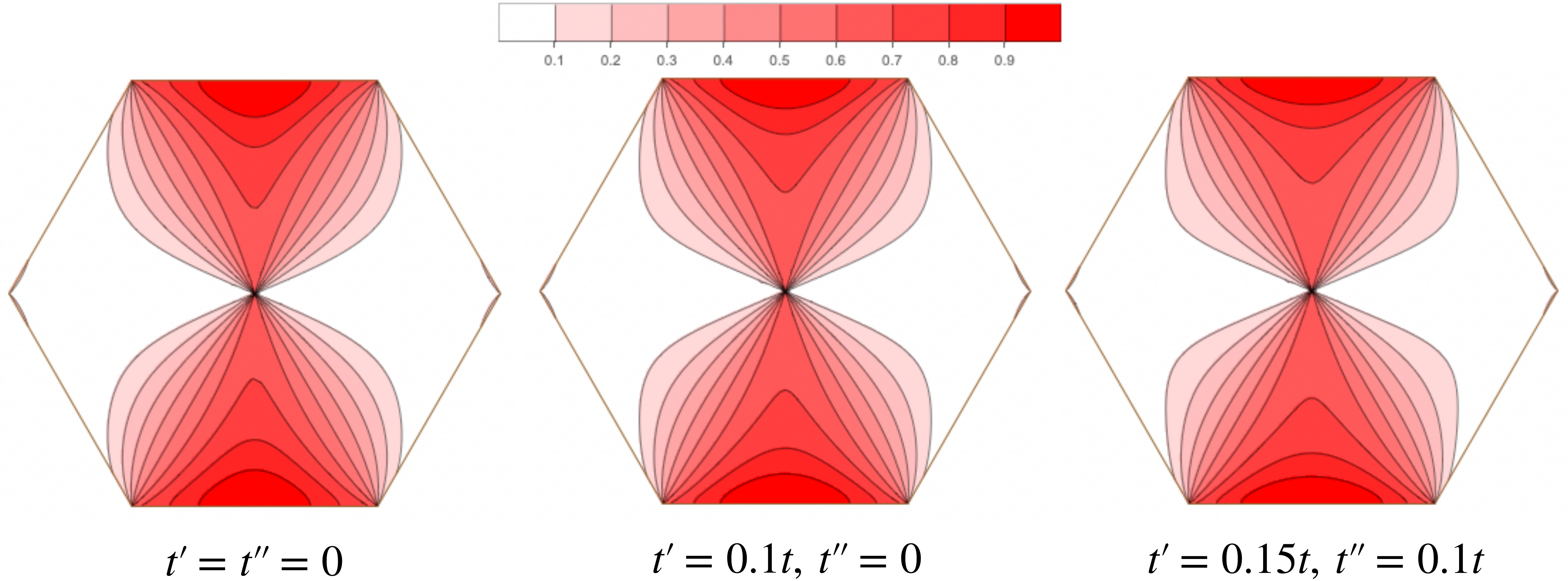}
	\caption{Momentum space configuration of $|u_{A}(\Bk)|^2$ for different cases. Here we set $t=1$.}\label{fig:SI_check}
\end{figure}
\section{Map from lattice interactions onto the patch model} 
\label{sec:}


For the three patch model, there are only four symmetry allowed interactions $g_1, g_2, g_3$ and $g_4$, which are defined through the most general interacting Hamiltonian
\begin{equation}
  H_I=g_1\psi^\dagger_a\psi^\dagger_b\psi_a\psi_b+g_2\psi^\dagger_a\psi^\dagger_b\psi_b\psi_a+g_3\psi^\dagger_a\psi^\dagger_a\psi_b\psi_b+g_4\psi^\dagger_a\psi^\dagger_a\psi_a\psi_a
\end{equation}
where $a,b\in(M_1,M_2,M_3)$ are the patch indices. When generalized to six patch model, we have 
\begin{equation}
H_I=\sum_{i,j=1}^4\sum_{\left\{ \alpha_j, \tau_j \right\}}g_{ij}\psi^\dagger_{\alpha_1\tau_1}\psi^\dagger_{\alpha_2\tau_2}\psi_{\alpha_3\tau_3}\psi_{\alpha_4\tau_4},
\end{equation} 

All the interactions can be projected onto the three patch interactions $g_i$ using Eq.\eqref{eq:interference}. For the onsite Hubbard $U$, we have 
\begin{equation}
  H_U=U\sum_s\sum_{k,k',q}u^*_s(k)u^*_s(k')u_s(k'-q)u_s(k+q)\psi^\dagger_\uparrow\psi^\dagger_\downarrow\psi_\downarrow\psi_\uparrow
\end{equation}
where the four fermion term $\psi^\dagger\psi^\dagger\psi\psi$ has the same momentum configurations as the $u^*u^*uu$ factors. Because all these $u_s$ factors have the same sublattice index, and according to \eqref{eq:interference}, all the four fermion operators must be in the same patch, otherwise the term vanishes. Therefore we have
\begin{equation}
  g_{4j}(0)=U
\end{equation}
where $g_{4j}(0)$ means the bare interaction, i.e. before RG.

For $H_V$ term, its corresponding form in momentum space is 
\begin{equation}
	H_V=2V\sum_{k,k',q}\cos(\bm{q}\cdot\bm{\alpha}_1)u_A^*(k)u_B^*(k')u_B(k'+q)u_A(k-q)\psi^\dagger_\sigma\psi^\dagger_{\sigma'}\psi_{\sigma'}\psi_\sigma+...
\end{equation}
where $...$ stands for similar contributions but with $AC$ and $BC$ combinations, and $\bm{\alpha}_1=(1,\sqrt{3})/4$ connects two adjacent $A$ and $B$ sites. Because of the SI condition Eq.\eqref{eq:interference}, both $\Bk$ and $\Bk-\bm{q}$ should be on the $\bm{M}_1$ patch, so $\bm{q}=0$ and $\cos(\bm{q}\cdot\bm{\alpha}_1)=1$. By choosing different patches, we can also have $\cos(\bm{q}\cdot\bm{\alpha}_1)=-1$. This interaction apparently contributes to $g_{2j}$, thus
\begin{equation}
	g_{21}(0)=g_{23}(0)=-2V, ~~g_{22}(0)=g_{24}(0)=2V
\end{equation}

If we include the site-to-site pair hopping term  $H_C=K_0\sum_{\braket{ij}}P_i^\dagger P_j$ interaction, it is completely parallel to get something like
\begin{equation}
  K_0\sum_{k,k',q}2\cos(\bm{q}\cdot\bm{\alpha}_1)u_A^*(k)u_A^*(-k+q)u_B(k')u_B(-k'+q)\psi^\dagger_\uparrow\psi^\dagger_\downarrow\psi_\downarrow\psi_\uparrow+...
\end{equation}
Again we use the sublattice interference condition in \eqref{eq:interference} to conclude that for this kind of interaction we must have $\bm{q}=2\bm{M}_1$ and it contributes to $g_{3j}$. For the coefficients $u_s(\bm{q})$ and the fermions $\psi(\bm{q})$, we must have the condition $2\bm{M}_1=0$. However, this is not the case for the function $\cos(q\cdot\alpha_1)$, because $\bm{\alpha}_1$ is half of the unit cell primitive vector, and we have $\cos(q\cdot\alpha_1)=-1$ instead. Similarly, we can have situations when $\cos(q\cdot\alpha_1)=1$, depending on the combination of patches. The resulting projections are
\begin{equation}
	g_{31}(0)=g_{32}(0)=2K_0, ~~g_{33}(0)=g_{34}(0)=-2K_0
\end{equation}

The bond pair hopping is a bit more complex. We only show the $AB$ bond for simplicity, but other bonds are easy to obtain. First of all, the momentum space representation of the pair operators are
\begin{equation}
	\begin{aligned}
		P_{AB}^\dagger(\bm{q})=\sum_{\Bk}\left(c_{A\uparrow}^\dagger(\Bk)c_{B\downarrow}^\dagger(-\Bk+\bm{q})-c_{A\downarrow}^\dagger(\Bk)c_{B\uparrow}^\dagger(-\Bk+\bm{q})\right)e^{i(\bm{q}/2-\Bk)\cdot\bm{\alpha}_1}\\
		P_{AB}(\bm{q})=\sum_{\Bk}\left(c_{B\downarrow}(-\Bk+\bm{q})c_{A\uparrow}(\Bk)-c_{B\uparrow}(-\Bk+\bm{q})c_{A\downarrow}(\Bk)\right)e^{-i(\bm{q}/2-\Bk)\cdot\bm{\alpha}_1}\\
		P_{BA}^\dagger(\bm{q})=\sum_{\Bk}\left(c_{B\uparrow}^\dagger(\Bk)c_{A\downarrow}^\dagger(-\Bk+\bm{q})-c_{B\downarrow}^\dagger(\Bk)c_{A\uparrow}^\dagger(-\Bk+\bm{q})\right)e^{i(\bm{q}/2-\Bk)\cdot\bm{\alpha}_1}\\
		P_{BA}(\bm{q})=\sum_{\Bk}\left(c_{A\downarrow}(-\Bk+\bm{q})c_{B\uparrow}(\Bk)-c_{A\uparrow}(-\Bk+\bm{q})c_{B\downarrow}(\Bk)\right)e^{-i(\bm{q}/2-\Bk)\cdot\bm{\alpha}_1}
	\end{aligned}
\end{equation}
Using this, we obtain for the on-bond pairing hopping,
\begin{equation}
	\begin{aligned}
		&J_0\sum_iP_{AB}^\dagger(\bm{r}_i)P_{AB}(\bm{r}_i)+(A\leftrightarrow B)=J_0\sum_qP_{AB}^\dagger(\bm{q})P_{AB}(\bm{q})+P_{BA}^\dagger(\bm{q})P_{BA}(\bm{q})\\
		&=2J_0\sum_{k,k',q}\sum_{\sigma\neq\sigma'}\cos[(\Bk-\Bk')\cdot\bm{\alpha}_1]u_A^*(k)u_B^*(-k+q)u_B(-k'+q)u_A(k')\psi^\dagger_\sigma\psi^\dagger_{\sigma'}\psi_{\sigma'}\psi_\sigma\\
		&+2J_0\sum_{k,k',q}\sum_{\sigma\neq\sigma'}\cos[(\Bk-\Bk')\cdot\bm{\alpha}_1]u_A^*(k)u_B^*(-k+q)u_A(k')u_B(-k'+q)\psi^\dagger_\sigma\psi^\dagger_{\sigma'}\psi_{\sigma'}\psi_\sigma
	\end{aligned}
\end{equation}
In the above expression, since both $\Bk$ and $\Bk'$ are in the $\bm{M}_1$ patch, $\Bk-\Bk'=0$ and the $\cos$ term equals $1$. It is the apparent that the first part of the result contributes to $g_{2j}$ while the second parts contributes to $g_{1j}$, with both positive $2J_0$. The same analysis can be applied to the nearest neighbor bond pair hopping, 
\begin{equation}
	\begin{aligned}
		&J_1\sum_iP_{AB}^\dagger(\bm{r}_i)P_{BA}(\bm{r}_i+\bm{\alpha})+P_{AB}^\dagger(\bm{r}_i)P_{BA}(\bm{r}_i-\bm{\alpha})+(A\leftrightarrow B)\\
		&=2J_1\sum_q\left(\cos\bm{q}\cdot\bm{\alpha}_1 P_{AB}^\dagger(\bm{q})P_{BA}(\bm{q})+A\leftrightarrow B\right)\\
		&=4J_1\sum_{k,k',q}\sum_{\sigma\neq\sigma'}\cos\bm{q}\cdot\bm{\alpha}_1\cos[(\Bk-\Bk')\cdot\bm{\alpha}_1]u^*_A(k)u^*_B(-k+q)u_B(k')u_A(-k'+q)\psi^\dagger_\sigma\psi^\dagger_{\sigma'}\psi_{\sigma'}\psi_\sigma\\
		&+4J_1\sum_{k,k',q}\sum_{\sigma\neq\sigma'}\cos\bm{q}\cdot\bm{\alpha}_1\cos[(\Bk-\Bk')\cdot\bm{\alpha}_1]u^*_A(k)u^*_B(-k+q)u_A(-k'+q)u_B(k')\psi^\dagger_\sigma\psi^\dagger_{\sigma'}\psi_{\sigma'}\psi_\sigma
	\end{aligned}
\end{equation}
Because of the SI condition, $\Bk$ is at $\bm{M}_1$ while $\Bk'$ is at $\bm{M}_2$, then $\Bk-\Bk'=\bm{M}_3$. It then follows that $\cos\bm{q}\cdot\bm{\alpha}_1\cos[(\Bk-\Bk')\cdot\bm{\alpha}_1]=-1$. Similar to the $J_0$ term, $J_1$ term also has two distinct parts, of which the first part contributes to $g_{2j}$ and the second part contributes to $g_{1j}$. 

\section{Renormalization group analysis} 
\label{sec:renormalization_group_analysis}


Having projected all the lattice interactions onto $g_{ij}$, we can investigate how they affected one another when high energy degrees of freedom is integrated out. This is captured by the following one-loop RG equations\cite{Isobe2018},
\begin{equation}
	\begin{aligned}
		\dot{g}_{11}&=2d_1(g_{11}g_{22}+g_{31}g_{32}-g_{13}^2-g_{33}^2),\\
		\dot{g}_{12}&=2d_1(g_{12}g_{24}+g_{32}g_{34}-g_{12}g_{14}-g_{32}g_{34}),\\
		\dot{g}_{13}&=2d_1(g_{13}g_{22}+g_{32}g_{33}-g_{11}g_{13}-g_{31}g_{33}),\\
		\dot{g}_{14}&=2d_1(g_{14}g_{24}-g_{14}^2),\\
		\dot{g}_{21}&=d_1(g_{21}g_{24}+g_{31}g_{34}),\\
		\dot{g}_{22}&=d_1(g_{22}^2+g_{32}^2),\\
		\dot{g}_{23}&=d_1(g_{22}g_{23}+g_{32}g_{33}),\\
		\dot{g}_{24}&=d_1(g_{24}^2+g_{34}^2),\\
		\dot{g}_{31}&=d_1(g_{21}g_{34}+g_{24}g_{31}+2g_{11}g_{32}+2g_{22}g_{31}-4g_{13}g_{33})-(g_{31}g_{42}+g_{32}g_{41}+g_{31}g_{32}),\\
		\dot{g}_{32}&=d_1(2g_{22}g_{32}+2g_{24}g_{32}-2g_{14}g_{32})-(2g_{32}g_{42}+g_{32}^2),\\
		\dot{g}_{33}&=d_1(3g_{22}g_{33}+g_{23}g_{32}+2g_{13}g_{32}-2g_{11}g_{33}-2g_{13}g_{31})-(g_{33}g_{44}+g_{34}g_{43}+g_{33}g_{34}),\\
		\dot{g}_{34}&=d_1(4g_{24}g_{34}-2g_{14}g_{34})-(2g_{34}g_{44}+g_{34}^2),\\
		\dot{g}_{41}&=
		-(g_{41}g_{42}+2g_{31}g_{32}),\\
		\dot{g}_{42}&=
		-(g_{42}^2+2g_{32}^2),\\
		\dot{g}_{43}&=
		-(g_{43}g_{44}+2g_{33}g_{34}),\\
		\dot{g}_{44}&=
		-(g_{44}^2+2g_{34}^2).
	\end{aligned}\label{eq:gijRG}
\end{equation}
where $\dot{g_i}=d g_i/dy$ and $y=\nu_0\ln^2(\Lambda/T)$ with $\Lambda$ the UV cutoff and $\nu_0$ the density of states. The nesting parameters are
\begin{equation}
	d_1=\frac{d\Pi_{ph}(\bm{M}_\alpha,T)}{dy}\approx \frac{\Pi_{ph}(\bm{M}_\alpha,T)}{\Pi_{pp}(0,T)}.
\end{equation}
Note that $\Pi_{ph}(\bm{M}_\alpha,T)$ contains $\ln^2$-divergence as well and  $d_1$ reaches the maximum value of $\frac{1}{2}$ in the perfect nesting case. From their definition we know that $d_1$ is related to S/CDW.

It is interesting to note that, in the absence of SI effect, such that all $g_{ij}$ are set by Hubbard $U$ initially, one can make take the condition that $g_{ij}=g_i$. Then Eqs.\eqref{eq:gijRG} reduces to the well know RG equations for the conventional 3-patch model,
\begin{equation}
	\begin{aligned}
		\dot{g}_1&=2d_1g_1(g_2-g_1),\\
		\dot{g}_2&=d_1(g_2^2+g_3^2),\\
		\dot{g}_3&=2d_1g_3(2g_2-g_1)-(2g_3g_4+g_3^2),\\
		\dot{g}_4&=
		-(g_4^2+2g_3^2).
	\end{aligned}
\end{equation}
consistent with the 3-patch generalization of Ref.\cite{Furukawa}, and also with Ref.\cite{PhysRevB.100.085136}.

In the weak coupling limit, all these $g_{ij}$ behaves in a similar manner close to the critical point $y_c$, namely,
\begin{equation}
	g_{ij}=\frac{G_{ij}}{y_c-y}.\label{eq:Gij}
\end{equation}
To identify the leading order in the weak coupling limit, we first consider the RG equations for various order parameters. We will consider charge-density-wave, spin-density-wave, uniform superconductivity and pair-density-wave. The renormalization for these vertices are shown in Fig.
For the CDW, the order parameter can be written as $\Delta_{\text{CDW}+}$ and $\Delta_{\text{CDW}-}$, where we use $+$ and $-$ sign to denote the parity of this order. The real and imaginary part of these CDW order parameters are governed by the following RG equations,
\begin{equation}
	\begin{aligned}
		\dot{\Delta}^{R}_{\text{CDW}+}=\frac{d_1}{2}(g_{22}-g_{33}-2g_{11}+g_{23}+g_{32}-2g_{31}-2g_{13}){\Delta}^R_{\text{CDW}+}\\
		\dot{\Delta}^R_{\text{CDW}-}=\frac{d_1}{2}(g_{22}-g_{33}-2g_{11}-g_{23}-g_{32}+2g_{31}+2g_{13}){\Delta}^R_{\text{CDW}+}\\
		\dot{\Delta}^I_{\text{CDW}+}=\frac{d_1}{2}(g_{22}+g_{33}-2g_{11}+g_{23}-g_{32}+2g_{31}-2g_{13}){\Delta}^I_{\text{CDW}+}\\
		\dot{\Delta}^I_{\text{CDW}-}=\frac{d_1}{2}(g_{22}+g_{33}-2g_{11}-g_{23}+g_{32}-2g_{31}+2g_{13}){\Delta}^I_{\text{CDW}+}\\
	\end{aligned}
\end{equation}
Similarly for SDW orders we have 
\begin{equation}
	\begin{aligned}
		\dot{\Delta}^{R}_{\text{SDW}+}=\frac{d_1}{2}(g_{22}+g_{33}+g_{23}+g_{32}){\Delta}^R_{\text{SDW}+}\\
		\dot{\Delta}^R_{\text{SDW}-}=\frac{d_1}{2}(g_{22}-g_{33}+g_{23}-g_{32}){\Delta}^R_{\text{SDW}+}\\
		\dot{\Delta}^I_{\text{SDW}+}=\frac{d_1}{2}(g_{22}+g_{33}-g_{23}-g_{32}){\Delta}^I_{\text{SDW}+}\\
		\dot{\Delta}^I_{\text{SDW}-}=\frac{d_1}{2}(g_{22}-g_{33}-g_{23}+g_{32}){\Delta}^I_{\text{SDW}+}\\
	\end{aligned}
\end{equation}
For the uniform superconducting order, we obtain $s$- and $f$-wave orders from the six-patch model. In addition, we also obtain two degenerate $d$-wave and two degenerate $p$-wave. The order parameters obey the following RG equations,
\begin{equation}
	\begin{aligned}
		\dot{\Delta}_{s\text{SC}}&=-\frac{1}{2}(2g_{31}+2g_{32}+g_{41}+g_{42}){\Delta}_{s\text{SC}}\\
		\dot{\Delta}_{f\text{SC}}&=-\frac{1}{2}(-2g_{31}+2g_{32}-g_{41}+g_{42}){\Delta}_{f\text{SC}}\\
		\dot{\Delta}_{p\text{SC}}&=-\frac{1}{2}(g_{31}-g_{32}-g_{41}+g_{42}){\Delta}_{p\text{SC}}\\
		\dot{\Delta}_{d\text{SC}}&=-\frac{1}{2}(-g_{31}-g_{32}+g_{41}+g_{42}){\Delta}_{d\text{SC}}\\
	\end{aligned}
\end{equation}

Note that using Eq.\eqref{eq:Gij} all the RG equations for the order parameters can be written in the form
\begin{equation}
	\dot{\Delta}=\Gamma(g_{ij})\Delta=-\beta(G_{ij})\Delta/(y_c-y).
\end{equation}
Solving this equation, one obtains,
\begin{equation}
	\Delta(y)\propto(y_c-y)^{\beta(G_{ij})}
\end{equation}
The susceptibility for the corresponding order parameters behaves as 
\begin{equation}
	\dot{\chi}=\tilde{d}|\Delta|^2
\end{equation}
where $\tilde{d}=1$ for uniform superconducting orders, $\tilde{d}=d_s$ for PDW  and $\tilde{d}=d$ for charge or spin density waves. It can be inferred directly that 
\begin{equation}
	\chi(y)\propto (y_c-y)^{\gamma(G_{ij})}
\end{equation}
and 
\begin{equation}
	\gamma(G_{ij})=2\beta(G_{ij})+1.
\end{equation}
The leading order with be the one which has the most negative $\gamma(G_{ij})$. Below are the explicit expressions for $\gamma$:
\begin{equation}
	\begin{aligned}
		\gamma_{\text{CDW}+}^R&=-d_1(G_{22}-G_{33}-2G_{11}+G_{23}+G_{32}-2G_{31}-2G_{13})+1\\
		\gamma_{\text{CDW}-}^R&=-d_1(G_{22}-G_{33}-2G_{11}-G_{23}-G_{32}+2G_{31}+2G_{13})+1\\
		\gamma_{\text{CDW}+}^I&=-d_1(G_{22}+G_{33}-2G_{11}+G_{23}-G_{32}+2G_{31}-2G_{13})+1\\
		\gamma_{\text{CDW}-}^I&=-d_1(G_{22}+G_{33}-2G_{11}-G_{23}+G_{32}-2G_{31}+2G_{13})+1\\
		\gamma_{\text{SDW}+}^R&=-d_1(G_{22}+G_{33}+G_{23}+G_{32})+1, ~~\gamma_{\text{SDW}-}^R=-d_1(G_{22}-G_{33}+G_{23}-G_{32})+1\\
		\gamma_{\text{SDW}+}^I&=-d_1(G_{22}+G_{33}-G_{23}-G_{32})+1,~~\gamma_{\text{SDW}-}^I=-d_1(G_{22}-G_{33}-G_{23}+G_{32})+1\\
		\gamma_{s\text{SC}}&=(2G_{31}+2G_{32}+G_{41}+G_{42})+1,~~\gamma_{f\text{SC}}=(-2G_{31}+2G_{32}-G_{41}+G_{42})+1\\
		\gamma_{p\text{SC}}&=(G_{31}-G_{32}-G_{41}+G_{42})+1, ~~\gamma_{d\text{SC}}=(-G_{31}-G_{32}+G_{41}+G_{42})+1\\
	\end{aligned}
\end{equation}
As a result, we need to first solve the RG equations for $g_{ij}$ to obtain $G_{ij}$, and then obtain $\gamma$'s from the above equation to identify the leading weak coupling instability. 

\begin{figure}
	\includegraphics[width=17cm]{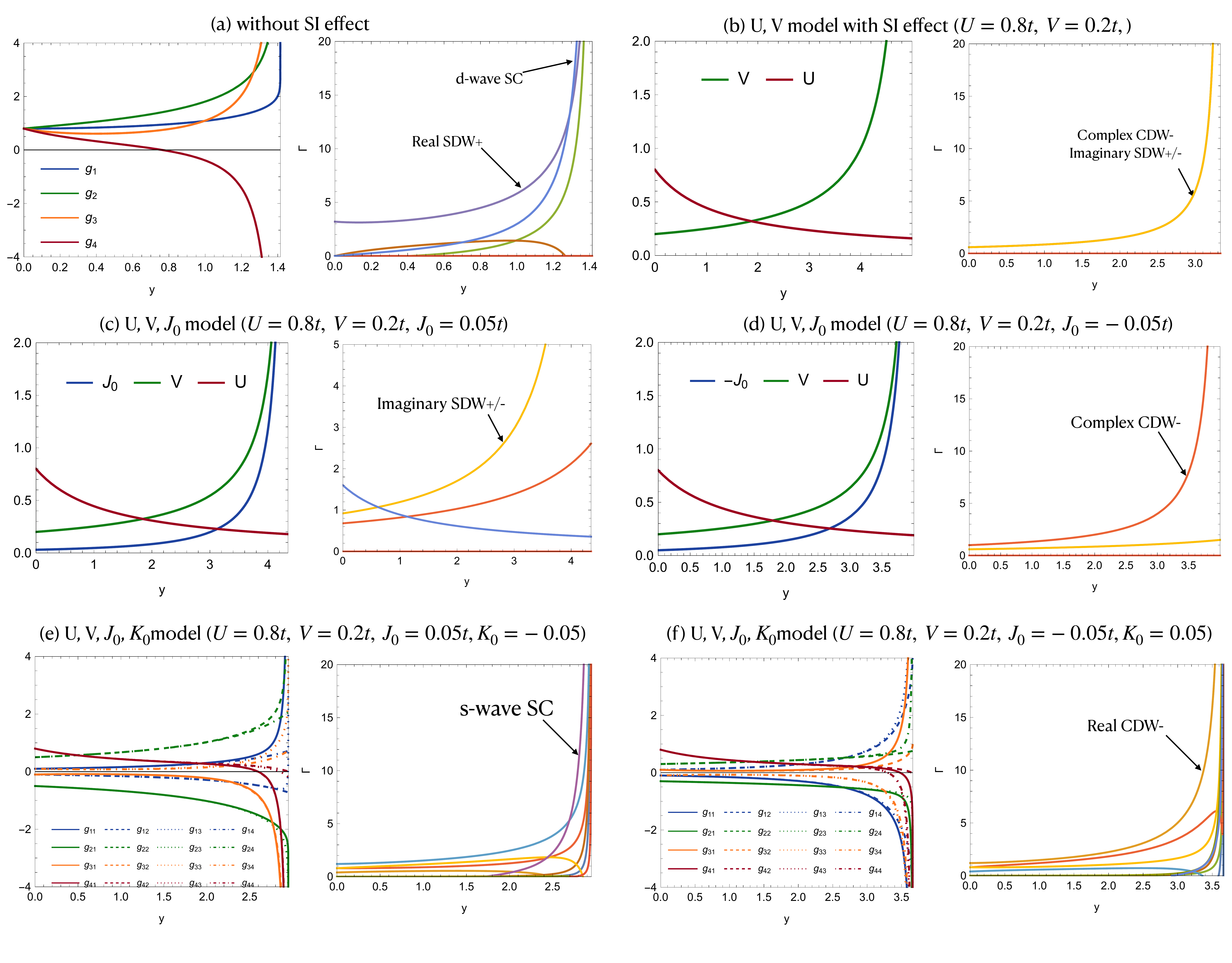}
	\caption{Solutions of the RG equation with finite interactions. (a), when there is no SI effect, then all the interactions are set by the largest $U$. In this case, all $g_{ij}$ are equal to $g_i$, and the RG result indicate the most divergent order is the $d$-wave SC order, consistent with the 3-patch model analysis for doped graphene in Ref.\cite{Nandkishore2012}. (b), with SI effect and when keeping only the largest $U$ and $V$ as initial input, the RG flows show that $U$ decays and $V$ grows. The resulting leading instability is degenerate complex CDW order with odd parity ($\Delta_{\text{CDW}-}$). (c) and (d), when the exchange interaction $J_0$ is further taken into account, the RG flows show that $J_0$ also grows, but is subleading to $V$. The leading instabilities in this situation is imaginary SDW with even and odd parity for $J_0>0$ and the complex CDW with odd parity for $J_0<0$. (e) and (f), when a finite initial $K_0$ is added into the RG flow, the RG results shows that near the critical energy scale, the presence of $K_0$ spoils the constant map between $g_{ij}$ and lattice interactions, but for $y<y_c$ the constant map is still a good approximation. Although we cannot identify the flow of each lattice interaction, we are still able to find the leading instabilities in weak coupling limit, which can be $s$-wave SC or real CDW order with odd parity.  }\label{fig:RGflow}
\end{figure}

In Fig.\ref{fig:RGflow} we show the weak coupling RG flows of the interactions for different situations, together with the flows of the order parameter vertex functions $\Gamma$ which help to illustrate the leading instability. 

We first show the results without the SI effect in Fig.\ref{fig:RGflow}(a), which can be realized at the m-type Van Hove filling for the Kagome lattice, or other hexagonal lattices such as triangular or honeycomb lattice. In these cases, all the interactions are set by the largest Hubbard $U$, and therefore $g_{ij}$ for all $j=1,..,4$ are identical. Then the resulting leading instability is the $d$-wave uniform superconductivity, consistent with the findings in Ref.\cite{Nandkishore2012}. The situation is quite different in the presence of SI effect. In Fig.\ref{fig:RGflow}(b) we show the results for $U$ and $V$ only model. In this particular case, we are able to find a constant map between the lattice interactions and the band fermion interactions, which clearly shows that as the energy scale is lowered, the $U$ decays and $V$ grows and we can have a low energy effective model where $V\gg U$. The resulting weak coupling instability is a complex CDW order with odd parity, and imaginary SDW order with both even and odd parity-- all of these orders are degenerate.

We next add some small but finite exchange interaction $J_0$ at the bare level and the results are shown in Fig.\ref{fig:RGflow}(c) and (d). If we only include $J_0$ but not $K_0$, we see that there is still a constant map between $g_{ij}$ and the lattice interactions, which clearly show that $U$ decays and both $V$ and $J_0$ increases, but $V$ still remains the largest. The weak coupling instabilities are imaginary CDW with both even and odd parity for $J_0>0$ and the complex CDW order with odd parity for $J_0<0$.

In Fig.\ref{fig:RGflow} (e) and (f) we show the RG results when all $U$, $V$, $J_0$ and $K_0$ are present. We see that the presence of a finite $K_0$ spoils the constant map between lattice interaction and $g_{ij}$ especially near the critical $y_c$. However, this destruction is minor at some intermediate energy scale and when the input $K_0$ is smaller. In this situation we can still have some approximate constant map between lattice interaction and $g_{ij}$, and from this we again can easily see that $V$ is the largest at some intermediate energy scale. In the weak coupling limit when the onset of instability is close to $y_c$, we can still extract the leading instability, for which we obtain $s$-wave uniform superconductivity with the parameters in Fig.\ref{fig:RGflow}(e) and real CDW order with odd parity with the parameters in Fig.\ref{fig:RGflow}(f). 

We emphasize that in all these RG results, the constant map between lattice interaction and $g_{ij}$ is present up to some intermediate energy scale, for which we have $V$ as the largest interaction. This observation makes us confident of using $V$-only model to study the competing orders in the strong coupling case.

\section{Calculation of the bare particle-hole and particle-particle bubble} 
\label{sec:calculation_of_the_bare_particle_hole_and_particle_particle_bubble}

Near the van Hove singularities at $\bm{M}_\alpha$ points, we can focus on the small patches around them. Here without loss of generality, we study the patches around $\bm{M}_A=(0,2\pi/\sqrt{3})$ and $-\bm{M}_{B}=(\pi,\pi/\sqrt{3})$. We denote the dispersions within these two patches as $\xi_1(\Bk)$ and $\xi_2(\Bk)$ respectively. The particle-hole bubble and the particle-particle bubble we will evaluate are given by 
\begin{equation}
	\begin{aligned}
	\Pi_{pp}^{(0)}&=T\sum_n\int \frac{d\Bk}{4\pi^2} \frac{1}{i\omega_n-\xi_1(\Bk)}\frac{1}{-i\omega_{n}-\xi_2(\Bk)}\\
		\Pi_{ph}^{(0)}&=T\sum_n\int \frac{d\Bk}{4\pi^2} \frac{1}{i\omega_n-\xi_1(\Bk)}\frac{1}{i\omega_{n}-\xi_2(\Bk)}\\
	\end{aligned}
\end{equation}
Near the vHS points, and in the model with only nearest neighbor hopping, we can approximate the dispersion as 
\begin{equation}
	\begin{aligned}
		\xi_1(\Bk)&=\frac{t}{2}(k_x^2-3k_y^2)=ta_+a_-,\\
		\xi_2(\Bk)&=-t(k_x^2+\sqrt{3}k_xk_y)=-ta_+(a_++a_-),
	\end{aligned}
\end{equation}
where $a_+=(k_x-\sqrt{3}k_y)/\sqrt{2}$ and $a_-=(k_x+\sqrt{3}k_y)/\sqrt{2}$
Using this, it is easy to see
\begin{equation}
	\begin{aligned}
		\Pi_{pp}^{(0)}&=\frac{-T}{4\sqrt{3}\pi^2t}\sum_n\int_{-\sqrt{\Lambda}}^{\sqrt{\Lambda}} da_+ da_- \frac{1}{i\omega_n-a_+a_-}\frac{1}{i\omega_n-a_+^2-a_+a_-}\\
		&=\frac{-T}{4\sqrt{3}\pi^2t}\sum_n\int_{-\sqrt{\Lambda}}^{\sqrt{\Lambda}}\frac{da_+}{a_+^3}\int_{-\sqrt{\Lambda}a_+}^{\sqrt{\Lambda}a_+}dx\left(\frac{1}{i\omega_n-x-a_+^2}-\frac{1}{i\omega_n-x}\right)
	\end{aligned}
\end{equation}
here we measure both $\omega_n$ and $T$ in units of $t$, and we have changed the variable as $x=a_+a_-$ in the second line. For the integration over $x$, we can shift the integration variables to get
\begin{equation}
	\begin{aligned}
		&\int_{-\sqrt{\Lambda}a_+}^{\sqrt{\Lambda}a_+}dx\left(\frac{1}{i\omega_n-x-a_+^2}-\frac{1}{i\omega_n-x}\right)\\
		=&\int_{-\sqrt{\Lambda}a_++a_+^2}^{\sqrt{\Lambda}a_++a_+^2}\frac{dx}{i\omega_n-x}-\int_{-\sqrt{\Lambda}a_+}^{\sqrt{\Lambda}a_+}\frac{dx}{i\omega_n-x}\\
		=&\int_{\sqrt{\Lambda}a_+}^{\sqrt{\Lambda}a_++a_+^2}\frac{dx(-x-i\omega_n)}{x^2+\omega_n^2}-\int_{-\sqrt{\Lambda}a_+}^{-\sqrt{\Lambda}a_++a_+^2}\frac{dx(-x-i\omega_n)}{x^2+\omega_n^2}\\
		=&-\int_{\sqrt{\Lambda}a_+-a_+^2}^{\sqrt{\Lambda}a_++a_+^2}\frac{dx x}{x^2+\omega_n^2}
	\end{aligned}
\end{equation}
Note we have used the fact that the imaginary part is odd in $\omega_n$ and will not survive after frequency summation, and is thus dropped.  
In low $T$ limit, we can replace $T\sum_n$ with $\int_{-\infty}^\infty\frac{d\omega_n}{2\pi}$ and carry out the frequency integration, this results in $\int d\omega_n 1/(x^2+\omega_n^2)=\pi /|x|$. Using this, it is easy to see
\begin{equation}
	\begin{aligned}
		\Pi_{pp}^{(0)}&=\frac{1}{4\sqrt{3}\pi^2t}\int_{-\sqrt{\Lambda}}^{\sqrt{\Lambda}}\frac{da_+}{a_+^3} a_+^2\text{sgn}(a_+)
	\end{aligned}
\end{equation}
Imposing an IR cutoff $\sqrt{T}$ for the integration, we obtain $\ln\Lambda/T$ for the result. Now if we assume that there is a small deviation from the van Hove filling, characterized by a nonzero $\mu$, we end up with the following expression, 
\begin{equation}
	\Pi_{pp}^{(0)}=\frac{1}{4\sqrt{3}\pi^2t}\ln\frac{\Lambda}{\max\{T,|\mu|\}}
\end{equation}
For the particle-hole bubble, similar analysis shows that 
\begin{equation}
	\Pi_{ph}^{(0)}=\frac{1}{8\sqrt{3}\pi^2t}\ln\frac{\Lambda}{\max\{T,|\mu|\}}\ln\frac{\Lambda}{\max\{T,\mu,|t'|\}}
\end{equation}
Comparing to $\Pi_{pp}^{(0)}$, there is an additional $\ln$ coming from the nesting of the Fermi surface in the particle-hole channel, which can be cut by either a finite shift from the van Hove filling ($\mu\neq0$), or including the next nearest neighbor hopping term $t'\neq0$.

\end{widetext}

\end{document}